\documentclass[12pt,a4paper]{article}

\usepackage{graphicx}
\usepackage{amsmath}
\usepackage{amssymb}
\usepackage{amsfonts}
\usepackage{epsfig}

\input{tcilatex}

\begin{document}

\title{Critical behavior of spin and polymer models with aperiodic 
interactions}
\author{T. A. S. Haddad and S. R. Salinas \\
Instituto de F\'{\i}sica\\
Universidade de S\~{a}o Paulo\\
Caixa Postal 66318\\
05315-970, S\~{a}o Paulo, SP, Brazil}
\maketitle

\begin{abstract}
We review and extend some recent investigations of the effects of aperiodic
interactions on the critical behavior of ferromagnetic $q$-state Potts
models. By considering suitable diamond or necklace hierarchical lattices,
and assuming a distribution of interactions according to a class of
two-letter substitution rules, the problem can be formulated in terms of
recursion relations in parameter space. The analysis of stability of the
fixed points leads to an exact criterion to gauge the relevance of geometric
fluctuations. For irrelevant fluctuations, the critical behavior remains
unchanged with respect to the uniform systems. For relevant fluctuations,
there appears a two-cycle of saddle-point character in parameter space. A
scaling analysis, supported by direct numerical thermodynamic calculations,
shows the existence of novel critical universality classes associated with
relevant geometric fluctuations. Also, we show that similar qualitative
results are displayed by a simple model of two directed polymers on a
diamond hierarchical structure with aperiodic bond interactions.

\end{abstract}

\section{Introduction}

In some recent publications\cite{phs1,phs2,phs3,mst}, we investigated the
critical behavior of a $q$-state ferromagnetic Potts model on a class of
hierarchical diamond lattices with aperiodic interactions. Taking advantage
of the lattice structure, and assuming a (layered) distribution of exchange
interactions according to suitable two-letter substitution rules, we were
able to write exact recursion relations in order to characterize the
critical behavior under the influence of geometric fluctuations. For
disordered ferromagnetic systems, the well-known Harris criterion predicts a
change in critical behavior if the underlying uniform system displays a
positive specific heat critical exponent. An extension of this heuristic
criterion has been proposed by Luck\cite{luck} in order to gauge the
relevance of geometric fluctuations associated with aperiodicity. We derived
an exact expression of an analog of Luck's criterion on a hierarchical
lattice. For weak fluctuations, there is a nontrivial symmetric fixed point
in parameter space that leads to the same critical behavior of the uniform
model. For strong enough fluctuations, this symmetric fixed point becomes
fully unstable, and there appears a cycle-two in parameter space, that is
shown to lead to a novel class of (aperiodic) critical behavior \cite
{phs2,phs3}. As the systems are quite simple, it has been feasible to
perform a direct thermodynamic analysis of the free energy to check the
predictions of scaling arguments.

Now we review and extend these calculations. In Section II, we show that the
same results are obtained for a $q$-state Potts model on a necklace
hierarchical lattice. Although the critical temperature may change, critical
exponents associated with either the symmetric fixed point or the cycle-two
attractor do display universal features that do not depend on the particular
choice of diamond or necklace hierarchical structures (and on particular
values of the interaction energies). In Section III, we consider a model of
two directed polymers on a diamond lattice\cite{mb1,mb2} with a (layered)
distribution of aperiodic interactions, according to some two-letter
substitution rules. The exact recursion relations in parameter space turn
out to be very simple, but with a similar structure as in the case of the
aperiodic Potts model. We then present some examples to show that, for
relevant geometric fluctuations, a symmetric fixed point becomes fully
unstable and there is again a novel cycle-two attractor of saddle-point
character in parameter space.

\section{Potts model with aperiodic interactions}

The successive application of the period-doubling two-letter rule, 
\begin{equation}
A\rightarrow AB,\qquad B\rightarrow AA,  \label{eq1}
\end{equation}
on an initial letter $A$, produces the sequence 
\begin{equation}
A\rightarrow AB\rightarrow ABAA\rightarrow ABAAABAB\rightarrow ....
\label{eq2}
\end{equation}
At each stage of this construction, the numbers $N_{A}$ and $N_{B}$, of
letters $A$ and $B$, can be obtained from the recursion relations 
\begin{equation}
\left( 
\begin{array}{c}
N_{A}^{\prime } \\ 
N_{B}^{\prime }
\end{array}
\right) =\mathbf{M}\left( 
\begin{array}{c}
N_{A} \\ 
N_{B}
\end{array}
\right) =\left( 
\begin{array}{cc}
1 & 2 \\ 
1 & 0
\end{array}
\right) \left( 
\begin{array}{c}
N_{A} \\ 
N_{B}
\end{array}
\right) ,  \label{eq3}
\end{equation}
where the eigenvalues, $\lambda _{1}=2$ and $\lambda _{2}=-1$, of the
substitution matrix $\mathbf{M}$ govern most of the geometric properties of
this sequence\cite{luckgod}. At order $n$, we can write 
\begin{equation}
N_{A}^{\left( n\right) }=\frac{2}{3}\lambda _{1}^{n}+\frac{1}{3}\lambda
_{2}^{n}\qquad \text{and\qquad }N_{B}^{\left( n\right) }=\frac{1}{3}\lambda
_{1}^{n}-\frac{1}{3}\lambda _{2}^{n},  \label{eq4}
\end{equation}
from which we have the asymptotic expressions $N_{A}^{\left( n\right) }\sim
\lambda _{1}^{n}$ and $\Delta N_{A}^{\left( n\right) }\sim \lambda _{2}^{n}$%
, for large values of $n$, where $\Delta N_{A}^{\left( n\right) }$ can be
regarded as a fluctuation around an asymptotic value. As the total number of
letters, at order $n$, is given by $N^{\left( n\right) }=N_{A}^{\left(
n\right) }+N_{B}^{\left( n\right) }$, we have $\Delta N_{A}^{\left( n\right)
}\sim \left( N^{\left( n\right) }\right) ^{\omega }$, with the wandering
exponent 
\begin{equation}
\omega =\frac{\ln \left| \lambda _{2}\right| }{\ln \lambda _{1}}.
\label{eq5}
\end{equation}

The ferromagnetic $q$-state Potts model is given by the Hamiltonian 
\begin{equation}
H=-q\sum_{\left( i,j\right) }J_{i,j}\delta _{\sigma _{i},\sigma _{j}},
\label{eq6}
\end{equation}
where $\sigma _{i}=1,2,...,q$, at all sites of a lattice, $J_{i,j}>0$, and
the sum refers to nearest-neighbor pairs of sites (for $q=2$, we regain the
standard Ising model). We now consider a Potts model on a hierarchical
necklace lattice, with $b=2$ bonds and $m$ branches, and assume that the
couplings can take just two values, $J_{A}$ and $J_{B}$, associated with a
sequence of letters produced by the period-doubling substitution. In Fig. 1,
we indicate some stages of this construction (for $b=2$ and $m=3$). Note
that the period-doubling sequence is perfectly suitable for a hierarchical
lattice with $b=2$ bonds. Also, note that the choice of the same
interactions along the branches turns out to mimic an aperiodic layered
structure in the corresponding Bravais lattice.

\begin{figure}
\begin{center}
\epsfbox{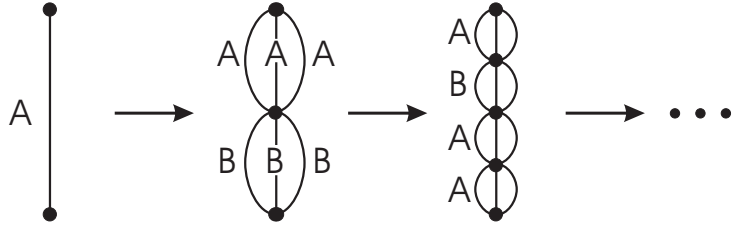}
\caption{Some generations of a hierarchical necklace lattice with 
$b=2$ bonds and $m=3$ branches. Letters $A$ and $B$ indicate the 
layered distribution of exchange interactions.}
\end{center}
\end{figure}

Now we decimate the internal degrees of freedom of the necklace structure
and write exact recursion relations for the reduced coupling parameters, 
\begin{equation}
x_{A}^{\prime }=\frac{x_{A}^{m}x_{B}^{m}+q-1}{x_{A}^{m}+x_{B}^{m}+q-2},
\label{eq7}
\end{equation}
and 
\begin{equation}
x_{B}^{\prime }=\frac{x_{A}^{2m}+q-1}{2x_{A}^{m}+q-2},  \label{eq8}
\end{equation}
where $x_{A,B}=\exp \left( \beta J_{A,B}\right) $ with $\beta =1/k_{B}T$.

In parameter space, besides the trivial fixed points, associated with zero
and infinite temperature, there is always a non-trivial symmetric fixed
point, associated with the critical behavior of the underlying uniform model
(see the sketch in Fig. 2a). The linearization of the recursion relations
about the symmetric fixed point $x^{\ast }$ leads to the matrix form 
\begin{equation}
\left( 
\begin{array}{c}
\Delta x_{A}^{\prime } \\ 
\Delta x_{B}^{\prime }
\end{array}
\right) =C\left( x^{\ast }\right) \widetilde{\mathbf{M}}\left( 
\begin{array}{c}
\Delta x_{A} \\ 
\Delta x_{B}
\end{array}
\right) ,  \label{eq9}
\end{equation}
where $C\left( x^{\ast }\right) $ is a structure factor and $\widetilde{%
\mathbf{M}}$ is the transpose of the substitution matrix. The eigenvalues of
this linear form are given by 
\begin{equation}
\Lambda _{1}=\lambda _{1}C\left( x^{\ast }\right) =2C\left( x^{\ast }\right)
,  \label{eq10}
\end{equation}
and 
\begin{equation}
\Lambda _{2}=\lambda _{2}C\left( x^{\ast }\right) =-C\left( x^{\ast }\right)
.  \label{eq11}
\end{equation}
It is easy to show that $\Lambda _{1}>1$. Note that we can write $\Lambda
_{1}=2^{y_{T}}$, with $y_{T}=D/\left( D-\alpha _{u}\right) $, where $D=\ln
\left( 2m\right) /\ln 2$ is the fractal dimension of the lattice and $\alpha
_{u}$ is the specific heat critical exponent of the underlying uniform
model. Also, it is easy to show that $\left| \Lambda _{2}\right| <1$, for $%
q<q_{c}$, and $\left| \Lambda _{2}\right| >1$, for $q>q_{c}$. In this last
case, for $q>q_{c}$, the symmetric fixed point becomes fully unstable, and
the geometric fluctuations are relevant (see the sketch in Fig. 2b). For $%
b=m=2$, both for necklace and diamond structures, we have $q_{c}=4+2\sqrt{2}%
=6.828427...$, which also corresponds to the critical number of states for
the relevance of disorder in a ferromagnetic Potts model on a fully
disordered diamond lattice \cite{dg}.

\begin{figure}
\begin{center}
\epsfbox{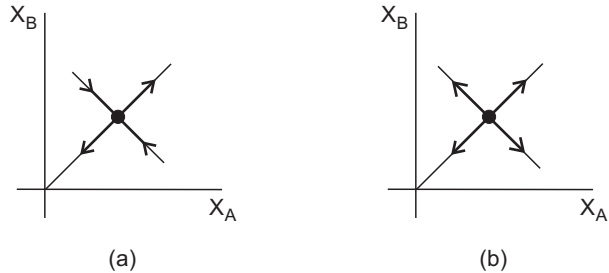}
\caption{Flow diagrams in the neighborhood of the symmetric fixed point.}
\end{center}
\end{figure}

For necklace and diamond lattices with $b$ bonds and $m$ branches, and a
suitable two-letter substitution sequence with period $\lambda _{1}=b$, we
can always write 
\begin{equation}
\Lambda _{1}=2C\left( x^{\ast }\right) =b^{D/\left( 2-\alpha _{u}\right) },
\label{eq12}
\end{equation}
where $D=\ln \left( mb\right) /\ln b$ is the fractal dimension of the
lattice, $\alpha _{u}$ is the specific heat critical exponent of the
underlying uniform model, and 
\begin{equation}
\Lambda _{2}=\lambda _{2}C\left( x^{\ast }\right) =\lambda _{1}^{\omega
}C\left( x^{\ast }\right) =b^{\omega }C\left( x^{\ast }\right) ,
\label{eq13}
\end{equation}
where the wandering exponent $\omega $ comes from Eq. (\ref{eq5}). We then
have 
\begin{equation}
\Lambda _{2}=b^{\omega -1+D/\left( 2-\alpha _{u}\right) },  \label{eq14}
\end{equation}
that leads to an exact version of Luck's criterion\cite{luck} for the
relevance of geometric fluctuations, 
\begin{equation}
\omega >1-\frac{D}{2-\alpha _{u}}.  \label{eq15}
\end{equation}
For the $q$-state Potts model on both diamond and necklace lattices with $%
b=2 $ bonds and $m=2$ branches, this inequality is equivalent to $q>q_{c}=4+2%
\sqrt{2}$, as we have pointed in the last paragraph. In this special case,
it also corresponds to $\alpha _{u}>0$, which is in agreement with the usual
form of Harris criterion for the relevance of disorder\cite{dg}. It should
be pointed out that, according to the usual expectations, given $q$, $b$ and 
$m$, the eigenvalues $\Lambda _{1}$ and $\Lambda _{2}$ assume the same
values for both diamond and necklace structures.

We now turn to the investigation of the critical behavior when the symmetric
fixed point is fully unstable. Numerically, we can see that, instead of a
hypothetical aperiodic fixed point, there appears a two-cycle, with a
saddle-point character, in parameter space. We then invoke scaling arguments
to write 
\begin{equation}
f\left( x\right) =g\left( x\right) +\frac{1}{b^{2D}}f\left( x^{\prime \prime
}\right) ,  \label{eq16}
\end{equation}
where $f\left( x\right) $ is the reduced free energy per bond, $g\left(
x\right) $ is a regular function, and $x^{\prime \prime }$ is a second
iterate of the recursion relations. Note that the factor $b^{2D}$ has to be
included because we need two iterates to go back to the vicinity of the
initial point in parameter space. This equation leads to the asymptotic
solution\cite{derrida} 
\begin{equation}
f\left( x\right) \sim \left| x-x^{\ast }\right| ^{2-\alpha }P\left( \frac{%
\ln \left| x-x^{\ast }\right| }{\ln \Lambda _{cyc}}\right) ,  \label{eq17}
\end{equation}
where $x^{\ast }$ is one of the points of the two-cycle, $\Lambda _{cyc}$ is
the largest eigenvalue of the linearization of the second iterate of the
recursion relations about any one of the points of the cycle, $P\left(
z\right) $ is an arbitrary function of period $1$, and the specific heat
critical exponent is given by 
\begin{equation}
\alpha =2-2\frac{\ln b^{D}}{\ln \Lambda _{cyc}}=2-2\frac{\ln \left(
mb\right) }{\ln \Lambda _{cyc}}.  \label{eq18}
\end{equation}

Let us show some calculations for the $q$-state Potts model on a necklace
lattice, with $b=2$ bonds, $m$ branches, and a distribution of exchange
interactions according to the period-doubling two-letter sequence: (i) for $%
q=7$ and $m=2$, the two-cycle is located at $\left( x_{A}^{\ast
},x_{B}^{\ast }\right) =\left( 2.299...,2.764...\right) $ and $\left(
2.587...,2.179...\right) $, with eigenvalues of the second iterate given by $%
\Lambda _{1}=3.993...\,$and $\Lambda _{2}=0.985...$. From Eq. (\ref{eq18}),
we have $\alpha =-0.00226...$, to be compared with the positive value $%
\alpha _{u}=0.010...$, for the uniform system; (ii) for $q=25$ and $m=2$,
the two-cycle is located at $\left( x_{A}^{\ast },x_{B}^{\ast }\right)
=\left( 2.634...,15.308...\right) $ and $\left( 6.246...,1.957...\right) $,
with eigenvalues of the second iterate given by $\Lambda _{1}=4.243...\,$and 
$\Lambda _{2}=0.343...$. From Eq. (\ref{eq18}), we have $\alpha =0.08174...$%
, to be compared with $\alpha _{u}=0.40456...$, for the corresponding
uniform system. For other values of $m$, we find qualitatively similar
results. The specific heat exponent is definitely depressed as compared to
the values for the underlying uniform system. Also, given $q$, $b$, and $m$,
the exponents assume the same values for diamond and necklace structures.

To check the validity of the scaling arguments, and the role of the
two-cycle as the responsible for the new critical behavior, we have
performed direct numerical analyses of the singularity of the thermodynamic
free-energy. For the $q$-state Potts model on the $b=2$ necklace
hierarchical lattice, and with the period-doubling rule, we can write the
free energy per bond as a series expansion, 
\begin{equation*}
f\left( x_{A},x_{B}\right) =-k_{B}T\sum_{n=0}^{\infty }\frac{1}{\left(
2m\right) ^{n}}\left\{ \frac{1}{3m}\ln \left[ \left( x_{A}^{\left( n\right)
}\right) ^{m}+\left( x_{B}^{\left( n\right) }\right) ^{m}+q-2\right] \right.
\end{equation*}
\begin{equation}
\left. +\frac{1}{6m}\ln \left[ 2\left( x_{A}^{\left( n\right) }\right)
^{m}+q-2\right] \right\} .  \label{eq19}
\end{equation}
All of our numerical checks fully confirm the scaling results. For example,
for $m=2$ and $q=100$, with $J_{A}/J_{B}=5$, a numerical analysis of the
specific heat divergence at the critical temperature leads to $\alpha
=0.27\pm 0.03$, to be compared with the scaling prediction from the
two-cycle, $\alpha =0.27204...$ (which is definitely different from the
value for the uniform system, $\alpha _{u}=0.64846...$). Within the error
estimates, we always obtain the same numerical results for all values of the
ratio $J_{A}/J_{B}$. Again, given $q$, $b$, and $m$, and for all values of
the ratio $J_{A}/J_{B}$, the exponents assume the same values for diamond
and necklace structures. From the numerical calculations, and particularly
for bigger values of $q$, we have been able to detect a log-periodic
oscillatory behavior of the thermodynamic functions. The period of
oscillation is compatible with Eq. (\ref{eq17}), which has also been
confirmed in the independent calculations of Andrade\cite{andrade}.

\section{Directed polymers with aperiodic interactions}

Along the lines of a paper by Mukherji and Bhattacharjee\cite{mb1}, we now
consider a model of two interacting directed polymers on a diamond lattice
(see Fig. 3). The polymers start at one end of the lattice and meet at the
other end. There is an attractive interaction if a bond of the lattice is
shared by two polymers.

\begin{figure}
\begin{center}
\epsfbox{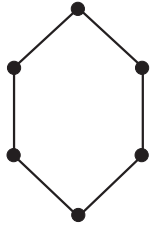}
\caption{Diamond hierarchical lattice with $b=3$ bonds  and $m=2$ branches.}
\end{center}
\end{figure}

Consider a diamond lattice ($b=2$ bonds and $m$ branches), with a layered
distribution of interactions, according to the two-letter period-doubling
sequence given by Eq. (\ref{eq1}). It is easy to write the recursion
relations 
\begin{equation}
y_{A}^{\prime }=\frac{1}{m}y_{A}y_{B}+\frac{m-1}{m},  \label{eq3.1}
\end{equation}
and 
\begin{equation}
y_{B}^{\prime }=\frac{1}{m}y_{A}^{2}+\frac{m-1}{m},  \label{eq3.2}
\end{equation}
where $y_{A,B}=\exp \left( \beta v_{A,B}\right) $, and $v_{A,B}>0$ is the
interaction energy at bonds of types $A$ and $B$, respectively. For $%
y_{A}=y_{B}=y$, we recover the recursion relation for the uniform model, 
\begin{equation}
y^{\prime }=\frac{1}{m}y^{2}+\frac{m-1}{m}.  \label{eq3.3}
\end{equation}
Besides the trivial fixed points, $y^{\ast }=1$ and $\infty $, associated
with zero and infinite temperatures, there is a nontrivial fixed point, $%
y^{\ast }=m-1$, which becomes physically acceptable, and is associated with
a binding-unbinding transition, for $m>2$ (there is no phase transition on
the simple diamond lattice with $m=2$ branches).

The recursion relations (\ref{eq3.1}) and (\ref{eq3.2}) are so simple, that
it easy to see that, for $m>2$, there is no physically acceptable nontrivial
fixed points except the symmetric fixed point $y_{A}^{\ast }=y_{B}^{\ast
}=y^{\ast }$ \ The linearization of the recursion relations in the
neighborhood of this symmetric fixed point leads to matrix form 
\begin{equation}
\left( 
\begin{array}{c}
\Delta y_{A}^{\prime } \\ 
\Delta y_{B}^{\prime }
\end{array}
\right) =\frac{y^{\ast }}{m}\left( 
\begin{array}{cc}
1 & 1 \\ 
2 & 0
\end{array}
\right) \left( 
\begin{array}{c}
\Delta y_{A} \\ 
\Delta y_{B}
\end{array}
\right) ,  \label{eq3.4}
\end{equation}
with eigenvalues $\Lambda _{1}=2y^{\ast }/m=2\left( m-1\right) /m$, and $%
\Lambda _{2}=-y^{\ast }/m=-(m-1)/m$. Therefore, if $m>2$, we have $\Lambda
_{1}>1$ and $\left| \Lambda _{2}\right| <1$, which shows that the geometric
fluctuations are completely irrelevant in this case.

We now turn to a more interesting case. Consider a diamond lattice with $b=3$
bonds and $m$ branches (in Fig. 3, we sketch a diamond lattice with $b=3$
bonds and $m=2$ branches). Suppose that the (layered) interactions are
chosen according to the period-$3$ two-letter sequence, $A\rightarrow ABB$
and $B\rightarrow AAA$. The substitution matrix is characterized by the
eigenvalues $\lambda _{1}=b=3$ and $\lambda _{2}=-2$, with the wandering
exponent $\omega =\ln 2/\ln 3=0.630092...$. The new recursion relations are
given by 
\begin{equation}
y_{A}^{\prime }=\frac{1}{m}y_{A}y_{B}^{2}+\frac{m-1}{m},  \label{eq3.5}
\end{equation}
and 
\begin{equation}
y_{B}^{\prime }=\frac{1}{m}y_{A}^{3}+\frac{m-1}{m}.  \label{eq3.6}
\end{equation}
There is just a single nontrivial fixed point, for $m>3$, at a symmetric
location, 
\begin{equation}
y_{A}^{\ast }=y_{B}^{\ast }=y^{\ast }=-\frac{1}{2}+\frac{1}{2}\sqrt{4m-3}.
\label{eq3.7}
\end{equation}
The linearization in the neighborhood of this fixed point leads to the
matrix form 
\begin{equation}
\left( 
\begin{array}{c}
\Delta y_{A}^{\prime } \\ 
\Delta y_{B}^{\prime }
\end{array}
\right) =\frac{\left( y^{\ast }\right) ^{2}}{m}\left( 
\begin{array}{cc}
1 & 2 \\ 
3 & 0
\end{array}
\right) \left( 
\begin{array}{c}
\Delta y_{A} \\ 
\Delta y_{B}
\end{array}
\right) ,  \label{eq3.8}
\end{equation}
with the eigenvalues 
\begin{equation}
\Lambda _{1}=3\frac{y^{\ast 2}}{m}=\frac{3}{2m}\left[ 2m-1-\sqrt{4m-3}\right]
,  \label{eq3.9}
\end{equation}
and 
\begin{equation}
\Lambda _{2}=-2\frac{y^{\ast 2}}{m}=-\frac{1}{m}\left[ 2m-1-\sqrt{4m-3}%
\right] .  \label{eq3.10}
\end{equation}
For $3<m<3+\sqrt{5}$, it is easy to show that $\Lambda _{1}>1$, and $\left|
\Lambda _{2}\right| <1$. As in the case of the simple diamond lattice with $%
b=2$ bonds, geometric fluctuations are irrelevant and the critical behavior
is identical to the uniform case. However, for $m>3+\sqrt{5}=5.236068...$,
we have $\left| \Lambda _{2}\right| >1$, and the symmetric fixed point
becomes fully unstable. For example, for $m=5$, we have $y_{A}^{\ast
}=y_{B}^{\ast }=y^{\ast }=1.561552...$, with eigenvalues $\Lambda
_{1}=2.140568...$ and $\Lambda _{2}=0.951363...<1$. For $m=6$, however, we
have $y_{A}^{\ast }=y_{B}^{\ast }=y^{\ast }=1.791287...$, with eigenvalues $%
\Lambda _{1}=2.573958...$ and $\Lambda _{2}=1.143981...>1$. As in the case
of the Potts model, there is then a cycle-two, of hyperbolic character, in
parameter space. It is easy to locate this cycle at $\left( y_{A}^{\ast
},y_{B}^{\ast }\right) =\left( 1.419001...,2.267305...\right) $ and $\left(
2.049103...,1.309541...\right) $, with eigenvalues of the second iterate
given by $\Lambda _{1}=2.624300...>1$ and $\Lambda _{2}=0.772598...<1$.

\section{Conclusions}

In conclusion, we have taken advantage of suitable diamond and necklace
hierarchical lattices, and used a distribution of (layered) interactions
according to a class of two-letter substitution rules, to perform an exact
analysis of the effects of geometric fluctuations on the critical behavior
of ferromagnetic $q$-state Potts models. We derived an exact expression for
an analog of Luck's criterion on a hierarchical lattice. Also, due to the
simplicity of the formulation, we have been able to show that the novel
critical behavior under the influence of relevant geometric fluctuations is
governed by a two-cycle attractor in parameter space. This new universality
class can be described by scaling arguments, that are fully supported by
direct thermodynamic calculations for the singularity of the free energy (as
usual, critical exponents do not depend on details of the systems, as the
relative strength of the interactions). It is not difficult to extend these
calculations for an Ising model in the presence of direct and staggered
fields\cite{ghs1}, to analyze the effects of aperiodicity on the tricritical
behavior in a mixed-spin Ising system\cite{ghs2}, and to consider some
noncommutative sequences\cite{sw1}

We have also reported some results for the critical behavior of a model of
two directed polymers on a diamond lattice with a (layered) distribution of
aperiodic interactions (according to suitable two-letter substitution
rules). The exact recursion relations in parameter space turn out to be very
simple, but with a similar structure as in the case of the aperiodic Potts
model. We have presented some examples to show that, for relevant geometric
fluctuations, a symmetric fixed point in parameter space becomes fully
unstable and there is again a novel cycle-two attractor, of saddle-point
character, that governs a new class of (aperiodic) critical behavior.

This work has been supported by the Brazilian agencies FAPESP and CNPq.

\end{document}